\documentclass{aastex}
\usepackage[onecolumn]{emulateapj5}

\def \degpoint {${.}\!\!^{\circ}\!$}

\def \eg{{e.g.,}\ }

\def \etal{{et~al.\ }}
\def \ie{{\it i.e.,}\ }

\def \degpoint {${.}\!\!^{\displaystyle \circ}\!$}

\shorttitle{Intracluster Stars in the Virgo Cluster}

\shortauthors{Durrell et al.}

\begin{document}

\slugcomment{\medskip Accepted for publication in the Astrophysical Journal}


{\title{Intracluster Red Giant Stars in the Virgo Cluster\altaffilmark{1}}

\altaffiltext{1}{Based on observations with the NASA/ESA Hubble Space
Telescope, obtained at the Space Telescope Science Institute, which is operated
by the Association of Universities for Research in Astronomy (AURA), Inc.,
under NASA contract NAS 5-26555.}

\author{Patrick R. Durrell, Robin Ciardullo}
\affil{Department of Astronomy \& Astrophysics, The Pennsylvania State University
\\ 525 Davey Lab, University Park, PA 16802 \\ {\it pdurrell@astro.psu.edu, rbc@astro.psu.edu}}

\medskip

\author{John J. Feldmeier}
\affil{Department of Astronomy, Case Western Reserve University \\10900 Euclid
Ave., Cleveland, OH 44106-1712\\ {\it johnf@eor.cwru.edu}}

\medskip

\author{George H. Jacoby}
\affil{WIYN Observatory
\\ P.O. Box 26732, Tucson, AZ 85726\\ {\it jacoby@noao.edu}}

\smallskip

\and

\author{Steinn Sigurdsson}
\affil{Department of Astronomy \& Astrophysics, The Pennsylvania State
University \\ 525 Davey Lab, University Park, PA 16802\\
{\it steinn@astro.psu.edu}}

\medskip

\begin{abstract}

We have used the WFPC2 camera of the {\sl Hubble Space Telescope\/} to obtain deep
F814W images of a blank field in the Virgo Cluster located $41\arcmin$
northwest of M87.  We perform star counts in that field, and in another Virgo
field observed by \citet{ftv98}, and show that, when compared to the Hubble
Deep Field North and South, the Virgo Cluster contains an excess of objects
with magnitudes $I \gtrsim 27$.  We attribute this excess to a population of
intracluster red-giant branch (IC-RGB) stars.  By modeling the luminosity
function of these stars, we show that the tip of the Virgo RGB is at $I_{TRGB}
\sim 27.31^{+0.27}_{-0.17}$ and that the cluster contains a small, but
significant, excess of stars that are up to $\sim 1$~mag brighter than this
tip.  If this luminous component is due entirely to stars on the asymptotic
giant branch (AGB), it implies an age for the population of $> 2$~Gyr; if
foreground RGB stars contribute to the luminous tail, then the derived age for
the stars is older still.  The luminosity function also suggests that most of
the intracluster stars are moderately metal-rich ($-0.8 \lesssim$ [Fe/H]
$\lesssim -0.2$), a result consistent with that expected from stars that have
been tidally stripped from intermediate luminosity galaxies.  Additionally, a
comparison with the planetary nebulae in our field also supports this view,
although the existence of a more metal-poor population (from stripped dwarfs)
cannot be ruled out. Our derived average surface brightness, $\mu_I =
27.9^{+0.3}_{-0.5}$ mag arcsec$^{-2}$ for Virgo's diffuse component suggests
that intracluster stars contribute 10\% to 20\% of the cluster's total $I$-band
luminosity.

\end{abstract}


\section{Introduction}

Intracluster stars (stars associated with galaxy cluster {\it potentials,}
rather than with any particular galaxy) provide an important clue towards the
understanding of the formation and evolution of galaxies and galaxy clusters.
N-body simulations show that this diffuse component can be
produced by a number of processes.  For instance, tidal interactions between
merging galaxies \citep{mil83,weil97,dub99}, between a galaxy and the cluster
potential \citep{mer84,dub99}, and between galaxies during high speed
encounters \citep[\ie `galaxy harassment';][]{moore96, moore98} can all liberate
stars \citep[and globular clusters, \eg][]{west95} into intracluster space.
Alternatively, a significant number of intracluster stars may be created early
on during a cluster's initial collapse \citep{mer84}.  By observing these stars
and determining their photometric and kinematic properties, we can therefore
learn about the workings of tidal-stripping, the distribution of dark matter
around galaxies, and the initial conditions of cluster formation.

Unfortunately, observational studies of intracluster light are very difficult
due to its very low surface brightnesses (typically $\mu_B \gtrsim 27$~mag
arcsec$^{-2}$, or less than 1\% of the background sky).  Consequently, though
the first detection of diffuse intracluster light was made a full half-century
ago \citep{zw51} and there have been numerous studies thereafter
\citep[\eg][]{oem73,mat77,mel77,tk77,uson91,vg94,bern95,gonz00}, there is
little agreement about the most basic data.  For example, even in the
well-observed Coma Cluster, measurements of the fraction of intracluster light
range from less than 25\% \citep{mel77} to $\sim 50\%$ \citep{bern95} of the
total cluster luminosity.

An obvious complement to measurements of the diffuse light in clusters is the
direct detection and measurement of individual intracluster stars. Although this
is only possible in nearby clusters (\eg Virgo, Fornax, Centaurus),
investigations of individual stars have the advantage of removing
many sources of error that typically complicate surface brightness measurements
(\ie contamination by low-surface brightness dwarf galaxies, scattered light
from foreground stars, flat-fielding errors, etc.).  Moreover, by studying
individual stars, one has the hope of determining the underlying population's
age, metallicity, and dynamical properties.

Because of their probative value, searches for intracluster stars have become
common in recent years.  In particular, a number of wide-field on-band/off-band
[O~III] $\lambda 5007$ imaging surveys for intracluster planetary nebulae (IPN)
have been conducted in fields of the Virgo and Fornax Clusters \citep{tw97,
men97, feld98, feld00}.  These studies have confirmed the existence of large
numbers of intracluster stars, and have produced evidence to suggest that many
of these stars are of moderate age and metallicity.

Although IPN are a powerful probe of intracluster starlight, they do have some
limitations.   Spectroscopy of Virgo IPN candidates by \citet{kud00} and
\citet{free00}, as well as a blank-field imaging survey by \citet{rbc01} have
shown that not all objects detected through narrow-band $\lambda 5007$ filters
are planetary nebulae -- about 20\% of the detections in Virgo appear to be
Ly$\alpha$ galaxies at $z = 3.13$.  This source of contamination produces an
ambiguity in the IPN analysis, which can only be broken via time-consuming
spectroscopy.  In addition, in order to determine the total amount of
intracluster light from IPN observations, one needs to know the production rate
of bright planetaries normalized to the bolometric luminosity of the stellar
population.  Observations demonstrate that this quantity varies by almost an
order of magnitude depending on the stellar population \citep{peim90,rbc95};
this creates a fundamental problem of the interpretation the IPN observations.
Finally, planetary nebulae are relatively rare objects: typically it takes
$\sim 5 \times 10^8 L_\odot$ of stars to produce $\sim 1$ [O~III] bright
planetary.

An alternative approach to studying intracluster starlight is to search for the
constituent red giant (RGB) and asymptotic giant branch (AGB) stars of the
stellar population.  RGB stars are much more numerous than planetary nebulae,
and therefore surveys for their presence do not require wide-field telescopes.
Moreover, translating the number counts of red giants to total population
luminosity is much more straightforward: there is no ambiguity as with
PN production.  Finally, because the absolute magnitude of the red giant
branch tip is a function of metallicity (for relatively high metallicity
populations), the luminosity function of RGB and AGB stars allows us to
constrain both the metallicity and age of the stellar population.

\citet{ftv98} (hereafter FTV) were the first to detect individual RGB stars in
intracluster space.  By using the WFPC2 camera of the {\sl Hubble Space
Telescope\/} to take deep F814W ($I$-band) images of a ``blank'' field located
45$^\prime$ E of the central Virgo cluster galaxy M87, Ferguson \etal were able
to detect the presence of intracluster red giants through the statistical
excess of point sources over that seen in the Hubble Deep Field North. From
their data, Ferguson \etal concluded that intracluster stars make up $\sim
10\%$ of the total stellar mass of the system.

Unfortunately, the Ferguson \etal analysis was necessarily limited.  Because
their survey area consisted of only one WFPC2 field, Ferguson \etal found only
a small number of stars, and thus could not place significant constraints on
the age or metallicity of the intracluster population.  Moreover, with only the
one field, Ferguson \etal could not address the question of the overall
distribution of these stars.  Recent discoveries of low-surface brightness arcs
in other nearby clusters \citep{gw98,tm98,cr00} and large field-to-field
variations in the number density of Virgo IPN \citep{feld98} have demonstrated
that intracluster stars are not distributed uniformly. Consequently, to
constrain the underlying population of these stars and learn about their
large-scale distribution, additional fields must be studied.

Here we present the results of a second study of Virgo intracluster RGB stars.
We begin by describing our deep {\sl HST\/} observations of a Virgo blank field
located between M87 and M86, near the center of the cluster's ``sub-clump A''
\citep{bts87}.  We detail our reduction techniques, our photometric procedures,
and the artificial star simulations needed to measure the errors and
incompleteness of our measurements.  In section 3, we combine our data with
those of the FTV survey, and compare the raw Virgo Cluster stellar luminosity
function (LF) with that of the North and South Hubble Deep Fields.  We show
that Virgo possesses a significant excess of point sources that is due to the
cluster's population of RGB and (possibly) AGB stars.  In section 4, we model
the point-source luminosity function, and place constraints on the cluster's
distance, and on the age and metallicity of its intracluster population(s).
Finally, we discuss these results and compare them with other measures of
intracluster stars.

\section{Observations}

On 5 and 6 May 2000, we conducted a search for intracluster stars by imaging a
Virgo Cluster ``blank field'' through the F814W filter of the WFPC2 camera of
the {\sl Hubble Space Telescope.}  The field, located at $\alpha_{2000} = 12^h
28^m 16^s$, $\delta_{2000} = +12^\circ 41^\prime 16^{\prime\prime}$, lies about
41$^\prime$ NW of M87, near the center of Virgo's subclump `A' \citep{bts87}.
As Figure~1 illustrates, the field is positioned far away from the bright
galaxies of the cluster, and within a region surveyed for planetary nebulae by
\citet{feld00}.

The total integration time for our survey was 33800~s; these data were broken
up into 13 2600~s exposures, which were dithered to facilitate the removal of
bad pixels and ``fill out'' the image point-spread-function (PSF{}).  To
supplement these data, we also re-reduced and analyzed the images taken by FTV
in their blank field survey of Virgo.  These images are very similar to ours:
they consist of 13 dithered WFPC2 F814W images (10 exposures of 2600~s plus 3
exposures of 2500~s) centered on a location $45^\prime$~E of M87. Finally, to
serve as a control, we used data from two other locations well away from any
galaxy or cluster: the Hubble Deep Field North \citep[HDF-N;][] {wil96} and
South \citep[HDF-S;][]{wil00}.   For both HDF fields, we extracted a subset of
images which, when combined, created a summed image with noise characteristics
similar to that of our Virgo Field.  For HDF-N, our control image was made from
13 frames ($11 \times 2600$~s $+$ $2 \times 2700$~s) totaling 34000~s; for the
HDF-S, our reference image was formed from using 14 frames ($8 \times 2400$~s,
$3\times 2300$~s, and single images of 2500s, 2700s and 2800s) with a total
exposure time of 34100~s.  For this investigation, we only considered the three
WF fields of the instrument; because of its small field-of-view, the PC chip
did not contain enough stars to warrant analysis.

\subsection{Data Reduction}

We reduced all four fields in exactly the same manner.  First, we corrected our
images for the small variation in pixel area \citep[see][]{hol95,stet98} across
the WF field-of-view using a pixel-area mask provided to us by Peter Stetson.
We then re-registered and averaged the images using the tasks within IRAF{}.
This resulted in a single deep $I$ image for each chip, with virtually all the
cosmic rays removed.  To keep the noise characteristics of the chips relatively
uniform, we then excluded those regions of each field that were not contained
on most of the component images.  Although this is a minor point for the two
Virgo fields and HDF-N, the relative shifts (and rotations) of some of the
images that went into building the HDF-S field were rather large.  As a result,
the useable area for this field was only 3.85 arcmin$^2$, compared to
4.69~arcmin$^{2}$ for HDF-N and 4.73~arcmin$^2$ for the two Virgo fields.

Once the images were combined, we equalized the photon noise for each field.
Because the Virgo Cluster is located within $\sim 15^\circ$ of the ecliptic
plane, the background zodiacal light in this area of the sky is slightly larger
than that present in the Hubble Deep Fields.  This results in a slightly
noisier background on these frames.  To compensate for this effect and make the
comparison between fields as fair as possible, we therefore added an additional
(Gaussian) component of noise ($\sigma = 0.4$~ADU) to the HDF frames. When this
was done, the noise in our control fields was virtually identical to that of
the cluster fields.

Photometry was carried out on each of the combined $I$ images using the
DAOPHOT~II $+$ ALLSTAR packages \citep{stet87, stet90, stet92}, with a
3.5$\sigma$ detection threshold.  Since both the cluster and HDF fields are
sparsely populated, only one DAOPHOT pass was required; secondary runs yielded
few new detections.  For the photometric measurements, we chose to use the
$I$-band WF PSFs that were exhaustively derived for the {\sl HST\/} Distance
Scale Key Project \citep[\eg][]{hill98,mould00}; these data were kindly
provided to us by Peter Stetson.  Creation of our own PSF was problematic, due
to the lack of bright, unsaturated stars in our fields.  Moreover, in the few
cases where suitable PSF stars were available, our derived PSF was not
significantly different from that found by the Key Project.

A major concern for any deep survey which uses a single filter is the
possibility of spurious identifications at the frame limit.  In particular,
with the relatively low detection threshold used here, it is common for
positive noise excursions to be flagged as objects.  To measure this effect, we
applied our DAOPHOT~II 3.5$\sigma$ detection algorithm to the {\it inverse\/}
of each image, and derived (in essence) the luminosity function of noise peaks.
The result from this experiment (adding data from all 4 fields) is shown in
Figure~2. Note that virtually all the detected noise spikes are fainter than
the limiting magnitude ($I_{lim}$) of the survey (see below for the definition
of $I_{lim}$).  Although some spurious detections {\it brighter\/} than
$I_{lim}$ do exist, these objects contribute less than 1\% of the total number
of counts between $I= 26.6$ and $27.4$.   Since this is significantly smaller
than the $\sqrt{N}$ statistics that dominate our analysis, we are confident
that false detections are not an important problem in our analysis.

The next step in our reduction was to remove non-stellar sources from our
object lists.  To do this, we used the DAOPHOT $\chi$ parameter \citep{stet87}
and $r_{-2}$ image moment \citep{kron80}; both are effective point-source image
discriminators \citep[\eg][]{stet87,har91,mcl95}. First, we determined the
critical values of $\chi$ and $r_{-2}$ from both a visual inspection of the CCD
frames, and from the results of artificial star experiments (see below).  Any
object whose image parameters exceeded either critical value was classified as
non-stellar and excluded from the analysis. To maintain consistency, the same
values of $\chi_{crit}$ (1.5) and $r_{-2,crit}$ (1.16) were used on each frame,
and in our artificial star simulations.  From the experiments, it is clear that
at magnitudes brighter than $I = 27$, our star/galaxy discrimination algorithm
works well; at these magnitudes, most galaxies are successfully rejected, and
very few stars are flagged as non-stellar.  At fainter levels, however, the
discrimination technique break down and few objects are rejected by the
procedure.

\subsection{Calibration}

All four fields were calibrated using the prescription given in \citet{hol95}.
We first converted the ALLSTAR PSF-based magnitudes to 0.5~arcsec aperture
magnitudes using measurements of 1 to 6 bright field stars on each WF chip; the
rms scatter in this procedure was typically 0.02-0.04~mag.  These aperture
magnitudes were then changed to standard $I$ magnitudes via the transformation
\begin{equation}
I = F814W - 0.063 (V-I) + 0.025 (V-I)^2 + z_I
\end{equation}
\citep{hol95}.  The zero points used were those derived by \citet{hill98};
these are for a gain of 7 and include the $\sim 0.05$~mag correction required
for long exposures.  At this time, an additional 0.016~mag offset was also
added in, to correct for differences in normalization between the Stetson
pixel-area mask and the Holtzman calibration \citep[cf.][]{stet98}. Finally,
as we have no color information, we assumed $(V-I)=2$ for all objects. This
value adequately represents the color of both our target RGB stars and a
substantial fraction of the background galaxies. Note that the error introduced
by our assumption of color ($\sim 0.03$~mag) is small, compared to the $\sim
0.2$~mag photometric errors expected for RGB stars in Virgo.

\subsection{Artificial Star Experiments}

Because the RGB stars of Virgo are expected to be at or near our detection
limit ($I \sim 27$) it is extremely important to have a complete understanding
of the data's photometric errors and our observational incompleteness.  To
measure these quantities, we took advantage of our knowledge of the WFPC2 PSF
and performed a series of artificial stars experiments.   For each chip of each
field, we added a total of 10000 artificial stars (in 100 simulations, with 100
stars per simulation), and re-reduced the frames in precisely the same manner
as described above -- a single pass of DAOPHOT II/ALLSTAR, followed by the
removal of non-stellar images using our image-classification algorithm.   By
limiting the number of artificial stars added in each run to less than 10\% of
the observed objects, we ensured that our simulations did not significantly
change the object density (and therefore the image crowding) of the fields.
Similarly, by giving our artificial star luminosity function a positive slope
($dN/dI = 0.7$ to roughly mimic the expected behavior of stellar objects in the
fields) we improved our statistics at the faintest magnitudes, where most of
the real objects lie.

Figure~3 displays the results from one of our simulations: that for the Virgo~A
field.  The upper panel of the plot displays $f(I)$, photometric completeness
function, which we define as the number of artificial stars of magnitude $I$
divided by the number of those stars recovered. Note that, due to photometric
errors, the measured magnitudes of the artificial stars will not necessarily be
those that were originally simulated.  Specifically, for an ensemble of stars
with input magnitude $I_{in}$, the measured magnitudes will be distributed with
a dispersion $\sigma(I)$ about a measured magnitude, $I_{out}$, which is offset
from $I_{in}$ by an amount $\Delta I  = I_{out} - I_{in}$.  The center and
bottom panels of Figure~3 display these two quantities.

The trends displayed in Figure~3 are representative of all our simulations. For
bright objects, $f(I)$ is independent of magnitude ($f(I) \approx 1$), and the
measured magnitudes are very nearly equal to the input magnitudes.  However, as
the stars become fainter, the photometric uncertainties become larger, and
incompleteness becomes important.  Moreover, for the very faintest objects,
the measured magnitudes become systematically brighter than the true
magnitudes. This is due to a simple selection effect:  objects with positive
noise spikes are detectable, while those with negative noise are not.

We can characterize the incompleteness function of any field via the expression
\begin{equation}
f(I) = {{1}\over{2}}\left( 1 - {{\beta(I - I_{lim})}\over{\sqrt{1 +
            \beta^2(I - I_{lim})^2}}} \right)
\end{equation}
\citep{fl95}, where $I_{lim}$ is the limiting magnitude of the data
\citep[defined as the 50\% completeness level;][]{har90}, and $\beta$ is a
parameter that measures how steeply $f(I)$ declines from 1.0 to 0.0.  Table~1
combines the artificial star experiments for the three WF chips and gives the
most likely values of $I_{lim}$ and $\beta$ for each field.  The strong
similarity between the results for the Virgo fields and those for the control
fields demonstrates that our efforts to match the noise characteristics of each
field were successful.

\section{Luminosity Functions}
\subsection{The Background LFs : HDF-N $+$ HDF-S}

Because our star/galaxy discrimination algorithm breaks down near our survey
limit, the contribution of unresolved background galaxies to the faint end of
our stellar luminosity function is significant.  Consequently, before analyzing
the cluster luminosity functions, we must first examine the unresolved object
counts of our two control fields.

The top two panels of Figure~4 display the luminosity functions for the two
Hubble Deep Fields.  The solid points show the LFs of point sources; the
histogram gives the LFs of all objects.  Even a cursory inspection of the
figure reveals that our star/galaxy discriminator works well only at relatively
bright ($I \lesssim 27$) magnitudes; at fainter magnitudes it becomes
impossible to exclude galaxies based on their image parameters.  Moreover, the
data clearly show that, once the HDF-S counts are scaled to match the HDF-N
survey area, the LFs of the two fields are statistically indistinguishable.  We
can therefore sum the LFs to produce a composite `background' LF based on a
total area of 8.54 arcmin$^2$.  This control field LF is displayed in the
bottom panel of the figure.

It is important to note that the LFs of Figure~4 have {\it not\/} been
corrected for photometric completeness. Our tests on the inverse science images
(see section 2.2 above) demonstrate that a non-negligible fraction of the
counts fainter than $I_{lim}$ is due to noise. Consequently, measurements of
objects fainter than this limit are not reliable, and have been excluded from
our analysis. (However, stars {\it intrinsically\/} fainter than $I_{lim}$ can
be measured to be brighter than $I_{lim}$; these are accounted for in the
models below.)

\subsection{Virgo Cluster Fields}

Figure~5 displays the LFs for our Virgo~A field and for the FTV Virgo field.
Again, the solid points show the LFs for the point-like sources, while the
histogram illustrates the LF of all objects.  Figure~6 (and Table 2) shows the
stellar LFs with the background of Figure~4 removed.  The figures clearly
demonstrate that both Virgo fields contain a large excess of stellar objects at
magnitudes $I \gtrsim 27$, and a slight excess of objects at magnitudes $26.4
\lesssim I \lesssim 27$.  These are exactly the magnitudes where we would
expect to find Virgo Cluster RGB and AGB stars.

According to `galaxy harrassment' models \citep[eg.][]{moore99}, stars that are
removed from their parent galaxies via tidal encounters retain some memory of
the event which ejected them \citep[see also][]{john99}.  Consequently, if most
intracluster stars are formed in this manner, we might expect the distribution
of intracluster stars to be clumpy, or have significant substructure.  Since
our Virgo~A field and the FTV field are separated by more than a degree on the
sky, we might therefore expect the number of intracluster stars in the two
fields to be significantly different.  Figure~6 shows that this is indeed the
case; we detect more unresolved objects in our Virgo A field than in the FTV
field.  A simple count of objects brighter than $I_{lim}=27.65$ (to $I=26.55$)
shows that the Virgo~A field contains $1.48 \pm 0.17$ times more point sources
than the FTV field, \ie $618\pm 31$ stars {\it versus \/} $418\pm 28$.  These
raw star counts in themselves are not particularly meaningful, since they have
not been corrected for incompleteness.  However, the similar {\it shapes\/} of
the two background-corrected LFs is important: it, along with the derived RGB
tip magnitudes (see the next section) further strengthens interpretation that
the detected sources are the RGB stars of Virgo.

\section{Modelling the LFs}

The luminosity functions presented above are not corrected for incompleteness,
nor do they compensate for photometric uncertainties, which move objects from
one magnitude bin to another.  To derive the intrinsic LF of Virgo, we need to
correct for these effects.  The best way to do this is to create models of
Virgo Cluster's RGB + AGB population, convolve them with the photometric error
and incompleteness functions of Figure~3, compare the results to the observed
data, and identify the best fitting model.  Such a procedure not only
allows us to fit the observed data, but also enables us to derive the
density of unresolved stars in Virgo's intracluster space.

To represent Virgo's RGB population, we used a power-law luminosity function
with a slope of $d \log N / dI = 0.4$; this is based on the stellar
evolutionary tracks of \citet{gir00} for populations older than 4~Gyr and
sub-solar metallicities.  (The precise slope of the power law makes very little
difference to the final result.)  We truncated the bright end of the luminosity
at magnitude $I_{TRGB}$, a free parameter that represents the tip of the red
giant branch; we followed the faint end of the LF to $I_{cut} = 28.1$, where
the completeness function of Figure~3 drops below $f(I)=0.20$.  (Tests show
that fainter stars have little effect on the analysis.)  For the AGB component,
we again followed the models of \citet{gir00} and adopted a flat luminosity
function ($d \log N/dI = 0$) from $I_{TRGB}$ to $I_{TAGB} < I_{TRGB}$, with a
normalization such that the AGB component contributes 15\% of the stars at the
RGB tip.  For simplicity, we assumed that all the intracluster stars of Virgo
are at a single distance (\ie no depth effects are included).  We will comment
on the possibility of foreground RGB/AGB stars later.

Our RGB $+$ AGB luminosity function is shown schematically in Figure~7. Under
this formulation, our input LF has three free parameters: $I_{TRGB}$, $\Delta
I_{AGB} = I_{TRGB} - I_{TAGB}$, and the overall normalization of the model. The
most-likely values for these parameters are those that minimize $\chi^2$ in the
range $26.0 \le I \le 27.6$. Note that because the photometric errors are large
and the convolution function asymmetric, the observed luminosity function is
much smoother and slightly shifted towards brighter magnitudes than the input
LF{}. (This effect is most noticeable near $I_{lim}$.)  Note also that though
our limiting magnitude is $I\sim 27.65$, stars intrinsically fainter than this
contribute significantly to the observed ($I < I_{lim}$) luminosity function.
This is why an understanding of the photometric properties of these extremely
faint stars is crucial for the interpretation of an observed luminosity
function.

Table~3 details the parameters which best fit the LFs of Figure~6; in each
case, the minimized $\chi^2_{\nu}$ values are $\sim 1$, showing that our
RGB+AGB model is an acceptable match to the luminosity function.  The $1\sigma$
errors are derived from fits with $\chi_{\nu}^2 = \chi^2_{\nu,min} + 1.15$
\citep[\eg][]{numrec}, which is the appropriate value to use for a
three-parameter fit to 17 points.  As the table shows, the models for the
individual fields are not particularly well constrained: there are simply not
enough stars in the FTV or Virgo~A fields for a precise determination of the
stellar luminosity function.  However, since the shapes of the two observed
luminosity functions are similar, we can combine the datasets and improve our
measurement of the clusters' RGB and AGB population.  When we do this, the best
fit for the RGB tip in Virgo becomes $I_{TRGB} = 27.31^{+0.27}_{-0.17}$, and
$\Delta I_{AGB} = 0.8^{+0.2}_{-0.2}$. This model is plotted in the bottom panel
of Figure~6.  The improved precision of the model allows us to place
constraints on the age, metallicity, and importance of Virgo's intracluster
stars.

\subsection{Surface Brightness}

Our RGB and AGB detections allow us to estimate the $I$-band surface brightness
of Virgo's intracluster space.   We start with only the RGB stars: if we take
our best-fitting model of Virgo's intrinsic LF, and sum up the luminosity of
all the RGB stars down to $I_{cut}=28.1$ in the 9.46~arcmin$^2$ area of the two
survey fields, the result is an average surface brightness of $\mu_I = 30.48
^{+0.07}_{-0.03}$~mag/arcsec$^2$ (error based on 1$\sigma$ error in
$I_{TRGB}$).   Next, we add in the AGB stars. The contribution of this
component is a bit more uncertain, due to the larger error bars on $\Delta
I_{AGB}$, but when included, the Virgo intracluster surface brightness due to
all stars brighter than $I=28.1$ becomes $30.28^{+0.11}_{-0.08}$ (see also
Table~3).

Finally, we must include the substantial contribution of stars fainter than the
cutoff of our model luminosity function.  To compute this number, we used the
theoretical luminosity functions of \citet{gir00}, which include main-sequence
stars (down to $M_I \sim +10$) as well as stars on the RGB, AGB, and horizontal
branch.  We investigated a grid of 15 models, with ages of 4, 8, and 12~Gyr,
and with metallicities of $Z$ = 0.0001, 0.0004, 0.001, 0.004, and 0.008 (here,
$Z_{\odot} = 0.019$).  Our choice of models was driven by observational
considerations. Populations younger than 4~Gyr were excluded, since these
produce stars that are more than a magnitude brighter than the RGB tip; such
objects are not seen in our data.  Similarly, based on the brightness of the
RGB stars, we did not consider stars with solar or super-solar metallicities
(see the discussion below).

Since we do not know {\it a priori\/} the distance of Virgo's intracluster
stars (and therefore the absolute magnitude of the RGB tip), we parameterized
our luminosity corrections in terms of $\Delta I_{TRGB}$, the difference
between the cutoff of our model luminosity function ($I = 28.1$) and the tip of
the giant branch.  For each Girardi \etal model, we removed the AGB component
and computed $F$, the fraction of the total light generated
by RGB stars within $\Delta I_{TRGB}$ of the giant branch tip.  We then used
this ratio to correct our measured surface brightness for the contribution of
unobserved stars ($\mu_{true} = \mu_{obs} + \Delta \mu_{ML}$, where $\Delta
\mu_{ML} = 2.5{\rm log} F$ is the `missing light' contribution).

Table~4 summarizes our results.  The listed values of $\mu_{ML}$ (at a given
$\Delta I_{TRGB}$) are the mean values (and associated errors) derived from the
15 model luminosity functions.  This missing light correction does not depend
strongly on the population's age or metallicity, provided the stars are
relatively old and not $\sim$ solar or super-solar metallicity. When we add
this unseen component  ($\Delta \mu_{ML}=-2.38^{+0.18}_{-0.40}$, based on our derived
values for $\Delta I_{RGB}$) to our previous estimate of surface brightness, we
get $\mu_I = 27.9^{+0.3}_{-0.5}$ mag~arcsec$^{-2}$ for the $I$-band surface
brightness of the Virgo intracluster stars.  The quoted errors on this number
come from the statistical errors of the RGB and AGB components and the
population uncertainties given in Table~4. The rather asymmetric errors are due
to the unknown distance of the cluster: the larger the distance to Virgo, the
larger the correction for unresolved sources, and the brighter the derived
value of $\mu_I$.   If we assume a Virgo Cluster distance of 16~Mpc
\citep[\eg][]{har98,fer00,ton01}, this surface brightness translates into an
average luminosity surface density of $\sigma_{L,I} = 0.13^{+0.08}_{-0.05}
L_{\odot}{\rm pc}^{-2}$.

Our average value\footnote{We have used the average results from both fields
for most of the remaining analyses; there is no {\it a priori\/} reason to
choose which field $\mu_I$ is more `representative' as they are at similar
projected distances from M87.} of $\mu_I = 27.9^{+0.3}_{-0.5}$
mag~arcsec$^{-2}$ for the surface brightness of Virgo's intracluster stars is
roughly three times the luminosity derived by FTV in their study.   Part of
this difference is due to the $\sim 50\%$ greater number of stars detected in
the Virgo A field:  the derived surface brightness of the Virgo~A field is
$\mu_I = 27.7$, while that for the FTV field is only $\mu_I = 28.1$.  The
remaining offset is largely due to the correction for unresolved stars. FTV
assumed that the observed RGB $+$ AGB stars produce 16\% of the total
intracluster light, while our analysis suggests that this percentage is
$11^{+2}_{-4}$\%. While we cannot fully explain the discrepancy, it is possible
that the different results arise from differences in the treatment of the
cluster's AGB stars. The \citet{gir00} models demonstrate that the contribution
of these stars to a population's total luminosity is a sensitive function of
metallicity. Consequently, if one relies on the models for this phase of
evolution, the uncertainty in the result can be substantial.  By combining the
data from the two {\sl HST\/} survey fields, we have been able to
observationally constrain the luminosity of the AGB component, and thereby
improve the correction for unresolved stars.

With the luminosity of Virgo's intracluster stars now fixed, we can next
compare this number to the amount of light contained within the cluster's
galaxies.  To do this, we used the galaxy catalog of \citet{vcc} to find the
total $B$-band magnitudes of all member galaxies brighter than the catalog's
limit of $B_T \sim 20$.  (This limit is adequate for our purpose, since the
contribution of fainter systems to Virgo's total luminosity is negligible.)  We
then converted these $B$-magnitudes to $I$, using the observed $B-I$ colors of
early-type galaxies \citep{goud94} and a mean $(B-I)$ color for the late-type
galaxies \citep{dejong94}.  Finally, with the $I$ magnitudes of all the
galaxies in hand, we computed the cluster's $I$-band galactic luminosity
density via a series of circular apertures centered on the giant elliptical
galaxy M87.  (\citet{bts87} have shown that the luminosity density of Virgo
actually peaks $\sim 1^\circ$ NW of M87.  However, measurements of the cluster
kinematics \citep{bin93}, x-ray luminosity \citep{boh94, sch99}, and
three-dimensional shape \citep{jcf90, wb00, fer00, ton01} demonstrate that this
offset is likely due to the contribution of the M84/M86 Group, which is falling
into Virgo from behind.)

Figure~8 details how Virgo's galactic luminosity density changes with distance
from M87.  As can be seen, the galaxy light shows the expected monotonic
decline with radius; only the ``plateau'' between 0\degpoint 5 and 1\degpoint
75, which is due to the contribution of the M84/M86 group, interrupts this
trend.  The figure also shows our measurement of the luminosity density of
intracluster stars.  If we assume that the ``plateau'' value of the galaxy
luminosity density ($\sigma_{L,I} = 0.73 \pm 0.01 L_{\odot}{\rm pc}^{-2}$) is
representative of the central regions of the Virgo cluster, then the
intracluster component contributes $15^{+7}_{-5}\%$ of the luminous matter of
the cluster.  This value is larger than the $\sim 10\%$ derived by FTV value:
though FTV derived only $\sim 1/3$ as much intracluster light as we do here,
they also considered only the early-type galaxies in their measurement of
Virgo's galactic luminosity.  Consequently, our final numbers are not that
different.

If we consider the rather extreme possibility that our value for Virgo's
intracluster light is constant with radius beyond $\sim 2^{\circ}$ from M87,
then the importance of the diffuse component would increase to about $\sim 30 -
35 \%$ of Virgo's total light.  However, we believe that such a large density
for Virgo's intracluster light is unlikely, since in other clusters, the
intracluster component is known to decrease with radius
\citep[\eg][]{tk77,bern95}.

\section{Nature of the IC-RGB population}

\subsubsection{IC stars or M87 cD Envelope?}

Our observations of Virgo's intracluster stars raise an interesting question:
are the stars seen in the Virgo~A and FTV fields part of M87's cD halo? Almost
certainly, this issue has more to do with terminology than with science
\citep[see][and references therein]{vg99}.  In rich clusters, the cD envelope
of the central galaxy can often be traced over many hundreds of kpc
\citep[cf.\null][]{uson91,sch94}; the stars in such an envelope are almost
certainly not bound to any one galaxy.  Moreover, direct evidence for the
intracluster nature of cD envelopes comes from the observations of NGC~1399,
the central galaxy of Fornax.  In this system, the stellar velocity dispersion
monotonically decreases with radius out to a galactocentric distance of $R \sim
5$~kpc \citep[\eg][]{sag00}.  Once past this point, however, the velocity
dispersion of the envelope rapidly increases, until, by $R \sim 13$~kpc, the
dispersion of the stars and globular clusters match that of the cluster's
galaxies \citep{arn94,gril94,kp99}.  The data demonstrate that the stars in
this cD envelope kinematically belong to the cluster, not the central galaxy.
The radial dependence of M87's globular cluster velocity dispersion suggests
that the same thing is true for Virgo \citep{cote01}.

If the core of Virgo is similar to that of Fornax, then our derived surface
brightness for the intracluster stars of the Virgo~A and the FTV fields may
very well be consistent with that expected from a smooth extrapolation of M87's
luminosity profile.   To test this hypothesis, we used the surface photometry
measurements of \citet{dvn78} and \citet{sch86}; both authors have traced M87's
luminosity profile out to a distance of $\sim 20\arcmin$, where it falls below
their detection threshold of $\mu_B \sim 29$.   If we extrapolate these
measurements to the positions of our Virgo fields ($41\arcmin$ NW for Virgo~A,
$45\arcmin$ E for the FTV field) and assume Virgo's diffuse light has a $B-I$
color of 1.9 \citep{ftv98}, then we derive a background $I$-band surface
brightness of $\mu_I \sim 30.3$ and $\mu_I \sim 31$~mag~arcsec$^{-2}$,
respectively.  These values are more than mag fainter than we
observe.\footnote{Note that although our Virgo~A field is only $\sim 34^\prime$
from M86, this galaxy probably does not contribute much to our star counts. M86
is $\sim 0.2$~mag more distant than M87 \citep{jcf90, ton01}, thus its
contribution to our counts should be minimal.  Moreover, the extrapolation of
the galaxy's luminosity profile \citep{caon90} produces a surface brightness
that is substantially smaller than what is derived seen.}  Thus, the existence
of an additional luminous component to the cluster is a possibility. However,
this conclusion is far from certain; had we used the surface photometry of
\citet{cd78} \citep[see also][]{hhm98} instead of that of \citet{dvn78} and
\citet{sch86}, we would have derived surface brightnesses $\sim 2$ magnitudes
{\it brighter\/} than observed.  Such a discrepancy is not unexpected, given
the uncertainties associated with the extrapolation of these photographic
surveys.

The extremely high contrast Virgo images presented by \citet{arn96} and
\citet{weil97} are equally ambiguous.  Both of these images show that
extremely low-surface brightness ($\mu_B =28$ mag arcsec$^{-2}$) structures do
exist around the bright galaxies (M86 and M87) of Virgo. However, our Virgo~A
field lies between these features, and any tidal stream that may exist in the
region is well below the detection threshold of their plates. As a result, the
question of whether the detected intracluster stars are part of smooth cD
envelope or a component of an irregular tidal stream cannot as yet be answered.

\subsubsection{The RGB Population}

The absolute magnitude of the tip of the giant branch is an extremely useful
probe for extragalactic astronomy.  For populations with metallicities
[Fe/H] $\lesssim -0.8$ and ages more than a few Gyr, the RGB tip is
remarkably constant, $M_{I,TRGB} = -4.1 \pm 0.1$; this makes the feature a
useful extragalactic distance indicator \citep[\eg][]{lee93, sak97, har98,
fer00, bell01}.  Alternatively, in more metal rich populations, the tip of the
RGB is a sensitive measure of metallicity, since line-blanketing in the
$I$-band suppresses the emergent flux.  As a result, if one already knows the
distance to a stellar population, then the magnitude of the RGB tip can be
used to probe metallicity.  We can use this property to place a constraint on
the population of Virgo's intracluster stars.

In order to do this, we first consider the galaxy VCC~1104, a nucleated dwarf
elliptical (dE,N) projected onto the core of the Virgo Cluster.
According to \citet{har98}, this galaxy has a well-populated red giant branch,
whose tip lies at $I_{TRGB} = 26.87 \pm 0.06$. (We have corrected the Harris et
al.~value by 0.05~mag for consistency with the \citet{hill98} zero points
adopted in this paper.) This is $0.44^{+0.27}_{-0.18}$~mag brighter than the
RGB tip of Virgo's intracluster stars.  There are two possible explanations for
this offset.

The first possibility is that the observed intracluster stars are at the same
distance as VCC~1104, but are more metal rich.  Since VCC~1104 is a dwarf
galaxy \citep[$M_V = -16.2$;][]{dur97}, we can assume that its population is
dominated by old, relatively metal-poor stars ([Fe/H]$\sim -1$; C\^{o}t\'{e}
\etal 2000).  In this case, $M_{I,TRGB}$ of the galaxy is $-4.1$, and the $\sim
0.4$~mag offset between the two RGB tips implies $M_{I,TRGB} =
-3.66^{+0.27}_{-0.18}$ for the intracluster star population.  We can translate
this into a metallicity in two ways.  First, we can use stellar evolution
models: according to \citet{van00} and \citet{gir00}, absolute TRGB magnitudes
between $I \sim-3.9$ and $-3.4$ correspond to metallicities between
[Fe/H] $\sim -0.8$ and $-0.3$. Alternatively, we can use observations of
Galactic globular clusters to derive an empirical TRGB metallicity calibration.
The metal-rich globular NGC~6553 has a metallicity of [Fe/H] $\sim -0.2$
\citep{rut97, guar98, cohen99} and an RGB tip at $M_{I,TRGB} \sim -3.3 \pm 0.2$
\citep{sagar99}; 47 Tucanae has [Fe/H] $\sim -0.7$ \citep{carr97} and
$M_{I,TRGB} \sim -4.05 \pm 0.10$.  [The 47 Tuc measurement is based on the
fiducial of Da Costa \& Armandroff 1990, with an uncertainty derived solely
from the range of distance moduli quoted by \citet{hess87,da90,grat97}
and \citet{kal98}.]  The results from both methods imply that, if Virgo's
intracluster stars are at the same distance as VCC~1104, then most of the
stars must have metallicities between $-0.8 \lesssim$ [Fe/H] $\lesssim -0.2$.

Alternatively, it is possible that the populations of VCC~1104 and Virgo's
intracluster stars are similar ([Fe/H] $\sim -1$), but that the dwarf galaxy is
foreground to the bulk of the cluster.  If this is the case, the
(luminosity-weighted) distance modulus of the intracluster stars would be
$(m-M)_I=31.4^{+0.4}_{-0.2}$, and VCC~1104 would lie in the foreground by $\sim
3.5$~Mpc.  This seems highly unlikely.  \citet{bts87} have shown that the dE,N
galaxies of Virgo are strongly concentrated around the denser regions of the
cluster, and are thus good tracers of the Virgo core regions.  In this respect,
VCC~1104 is not exceptional: it is projected only $43\arcmin$ from M87 (and
$9\arcmin$ from our field).  For this reason alone, we would not expect
VCC~1104 to be a foreground object.  In addition, the implied distance of
VCC~1104 \citep[$15.4 \pm 0.9$~Mpc;][]{har98} is in excellent agreement with
almost all recent distance estimates to the Virgo Cluster core \citep{rbc98,
fer00, ton01}. If VCC~1104 were moved significantly into the foreground, then a
revision would be required in the Cepheid, surface brightness fluctuation, and
planetary nebula distance scales. A more likely explanation is that VCC~1104
is, indeed, located near the center of Virgo, and that many of the intracluster
stars of Virgo have metallicities between one-fifth and three-fifths solar.

\subsubsection{The AGB Population}

Just as the RGB stars of Virgo can be used to place limits on the IC stars'
metallicity, the AGB component of the system can be used to constrain the
population's age.  For example, the fits listed in Table~3 exclude AGB
populations that extend more than a magnitude above the tip of the red giant
branch, \ie $\Delta I_{AGB} = 0.8^{+0.2}_{-0.2}$.  According to the
\citet{gir00} models, this measurement demands that the dominant intracluster
population be older than about 2~Gyrs, regardless of metallicity.  If the
intracluster stars of Virgo have their origin in star-forming galaxies, then
their ejection into the intracluster medium must have taken place some time
ago.   Of course, this does not rule out the possibility that the stars were
recently removed from early-type galaxies \citep[\eg][]{weil97,kor01}.

Unfortunately, our data do not allow us to place an upper limit on the age of
the intracluster stars.   Although the mean metallicity of Virgo's intracluster
stars is probably [Fe/H] $\sim -0.5$, it is entirely possible that some of
Virgo's intracluster stars are more metal-poor than this.  This would result in
a population of RGB stars with magnitudes $\sim 0.4$~mag brighter than the
apparent RGB tip and contaminate our AGB measurement.

A more important source of confusion comes from our assumption about the
two-dimensional nature of the cluster.  The analysis above assumed that
all the intracluster stars in the Virgo~A and FTV fields are at the same
distance.   This is probably not the case.  Arguments based on the
surface brightness fluctuations of elliptical galaxies \citep{wb00}, the
Tully-Fisher relation of spiral galaxies \citep{foy93} and the luminosity
function of intracluster planetary nebulae \citep{rbc98, feld98} all suggest
that the Virgo Cluster is elongated along our line-of-sight.  If this is
true, and a small population of foreground stars do exist, some of our
putative AGB stars must be foreground RGB objects.  In fact, a careful
inspection of Figure~6 suggests that a very small population of objects with
$I\sim 25.8$ (about 0.6 mag brighter than our observed AGB tip)
may be present in the data.  If so, then this feature could be due to a
population of foreground AGB stars.  (The alternative explanation, a
population of young AGB stars associated with the primary RGB component,
is unlikely given the small number of stars observed.)  Because
of the low statistical significance of the feature, we have not included it
in any of our models.  Nevertheless, it is suggestive that a small
population of foreground stars may be present in our fields.

In our two-dimensional model of Virgo, intracluster stars with $I \sim 26.5$
are objects near the tip of the AGB{}.  If this is incorrect and these stars
are actually red giants, then it would imply the existence of a population that
is 3 to 5~Mpc in front of the cluster center.  Although this number seems
large, it is not totally inconsistent with the cluster's projected diameter of
2 to 3~Mpc.\footnote{Note that this does not contradict our earlier assertion
about VCC~1104 -- as stated above, dE,N galaxies preferentially lie near the
cluster core.}  Moreover, if the stars at the apparent tip of the AGB are
actually foreground red giants, then the actual AGB tip might be significantly
fainter than estimated ($\Delta I_{AGB} \rightarrow 0$).  This would imply a
significantly older stellar population.  A color-magnitude diagram from the
next generation of space-based instruments would greatly assist in clearing up
the uncertainties in both the age and metallicity of the IC population.

\subsection{Comparison with Planetary Nebula Surveys}

The data of the Virgo~A and FTV fields suggest that $\sim 10\%$ to $\sim 20\%$
of the Virgo Cluster's {\it total\/} light comes from its intracluster
population.  This is slightly smaller than the values of 20\% to 40\% that have
been found in the IPN searches published to date \citep{tw97,men97,feld98}. One
reason for this is that the early IPN counts underestimated the presence of
redshifted [O~II] $\lambda 3727$ and Ly$\alpha$ galaxies in the sample. When
this component is (statistically) removed, the IPN numbers drop by $\sim 20\%$
\citep{rbc01}.  A second reason for the discrepancy is that $\alpha$, the ratio
of bright planetary nebulae to parent population luminosity, is uncertain by
more than a factor of $\sim 5$ \citep{rbc95}.  Without some estimate of
$\alpha$, IPN-based measurements of intracluster stars carry a substantial
uncertainty.  Because the Virgo~A and FTV fields have both been included in the
recent [O~III] $\lambda 5007$ IPN survey of \citet{feld00}, it is possible to
use our star counts to estimate $\alpha$ in the intracluster environment.

To do this, we must first translate our measurement of intracluster $I$-band
surface brightness into an estimate of bolometric luminosity surface density.
This requires adopting reasonable values for the population's distance modulus
($(m-M)_o = 31.0$), $V-I$ color \citep[$\sim 1.2$;][]{ton01}, and bolometric
correction \citep[$\sim -0.8$;][]{jcf90}.  Using these values, we obtain a
bolometric-luminosity surface density for the population of $\sigma_{L,bol} =
1200^{+700}_{-300} L_{\odot}$ arcsec$^{-2}$ for our Virgo~A field. (This value
is 20\% larger than that derived by averaging both fields.)

Next, we need an estimate for the number of bright IPN in our field.
\citet{feld98} \citep[see also][]{feld00} originally found 69~IPN candidates in
a 244~arcmin$^2$ region surrounding our Virgo~A field, and 7 candidate IPN in a
200~arcmin$^2$ region around the FTV field.  Unfortunately, these counts are
suspect: clouds and variable seeing contaminated the Virgo~A IPN list with
background galaxies \citep{kud00,rbc01} and false detections
\citep[see][]{kud00}, and the IPN observations of the FTV field are not very
deep.  Consequently, instead of using the data of \citet{feld00}, our IPN
number counts from a newer IPN survey of Virgo~A \citep[for more details,
see][]{feld02}.  In a 1098~arcmin$^2$ region, \citet{feld02} found 32~IPN
candidates down to a limiting magnitude of $m_{5007} = 26.8$.   If we take this
density, statistically subtract background contaminants \citep{rbc01}, assume a
standard form of the planetary nebula luminosity function \citep{rbc89}, and
normalize the IPN density to that of the underlying intracluster light, then we
obtain a value for the bolometric luminosity-specific PN number density of
$\alpha_{2.5} = 23^{+10}_{-12} \times 10^{-9}$~PN~$L_{\odot}^{-1}$.  The
uncertainty comes from the Poissonian errors on the IPN counts, our estimate of
the bolometric luminosity surface density, and an estimated 0.1 magnitude error
in the IPN completeness magnitude.

Although the uncertainty in $\alpha_{2.5}$ is still rather large, it is good
enough to place a constraint on the stellar population of Virgo's intracluster
space.  \citet{rbc95} has noted a relation between $\alpha_{2.5}$ and the
absolute luminosity ($M_B$) of the PN's parent galaxy. Most luminous E/S0
galaxies \citep[including several in the Virgo Cluster;][]{jcf90} have low
values of $\alpha_{2.5}$ ($\sim 5-10 \times 10^{-9}$ PN $L_{\odot}^{-1}$); less
luminous, bluer systems have values of $\alpha_{2.5}$ that are high (up to
$\sim 50 \times 10^{-9}$~PN~$L_{\odot}^{-1}$).  Although the cause of this
behavior is not fully understood, it is apparent that old, metal-rich systems
are less efficient at making [O~III]-bright planetaries than are
intermediate-age, intermediate-metallicity populations.

Our results show that $\alpha_{2.5}$ in Virgo's intracluster population is a
factor of two larger than that found in massive early-type galaxies.  It is
also a factor of $\sim 2$ larger than the value of $\alpha_{2.5}$ derived for
Galactic globular clusters \citep{jmfkh97}.  The number is, however, similar
similar to that commonly seen in sub-$L^*$ systems (\ie $M_B > -20.5$). This
suggests a possible origin for the stars.  In addition, the relatively high
value of $\alpha$ also brings the IPN-based intracluster light measurements
more into line with what we derive from the IC RGB stars \citep{feld98}.

Of course, the above calculations assume that the RGB surface densities are
directly comparable to the IPN surface densities. If there is any spatial
structure in the intracluster light, then the density of stars in our small
{\sl HST\/} field may not be representative of that sampled by the wider-field
IPN measurements.  In fact, there is some evidence \citep{feld00} to suggest
that Virgo's intracluster light is non-randomly distributed.  However, at the
present time, the data are too sparse to draw any useful conclusion.  More
ground-based and space-based data are required to address this issue.

\subsection{Origin of the IC starlight}

Most models for the origin of intracluster light involve the removal of stars
and globular clusters from cluster galaxies via tidal encounters, either with
the overall cluster potential \citep{mer84,dub99} or with other cluster
galaxies during close encounters \citep[\eg][see Moore, Quilis, \& Bower 2000
for a review]{mil83,mr84,moore96,moore98}.   In the models, streams of material
are liberated from low-mass (sub-$L^*$) or low-surface brightness galaxies
during each tidal encounter, decreasing the population of such objects
dramatically.  Each tidal feature is only visible for a time, but, over a
Hubble-time, the process can produce a sizeable population of intracluster
stars.

One can argue that the most-likely candidates for tidal disruption are small,
loosely-bound dwarf galaxies.  Indeed, the fan-shaped halo surrounding M87 is
probably the remains of such an object \citep{weil97}, and numerical
simulations of CDM universes predict that large numbers of dwarf galaxies
should have been formed in clusters \citep[\eg][]{wf91,kauf93,cole94,klyp99}.
Moreover, \citet{cote98} have suggested that that the {\it metal-poor\/}
globular clusters associated with giant ellipticals (such as M87) can be
explained via the merging or stripping of large numbers of low-luminosity
dwarfs.   It is thus tempting to conclude that most (or all) intracluster stars
have their origin inside dwarf galaxies.  However, normal spiral and elliptical
galaxies are also susceptible to tidal forces.  In fact, the large, low
surface-brightness plumes of material seen in Coma and Centaurus almost
certainly come from normal-sized galaxies \citep{tm98,gw98,cr00}.  Hence the
question:  where do the intracluster stars come from?

Let us first hypothesize that the intracluster stars of Virgo are, indeed, the
remains of tidally stripped dwarf galaxies. The total bolometric luminosity of
all galaxies within $2^\circ$ of M87 is $\sim 10^{12} L_{\odot}$ \citep{vcc};
in this same region, the total luminosity of all dwarf and Im galaxies is $\sim
8 \times 10^{10} L_{\odot}$.  Thus, dwarf galaxies account for $\sim 8\%$ of
the total galactic light.  If we now assume that the radial distribution of
intracluster light is similar to that of the galaxies, then the star counts in
our two {\sl HST\/} fields imply that the total amount of intracluster light in
the $2^\circ$ core of Virgo is $\sim 1.7 \times 10^{11} L_{\odot}$.  In other
words, if the intracluster stars come primarily from disrupted dwarfs, then the
original number of such galaxies must have been roughly {\it three times\/}
that observed today.  This is possible, though difficult to prove.  Moreover,
in their consideration of the cD galaxy NGC~1399, \citet{hil99} concluded that,
though the assimilation of dwarfs could explain the luminosity of the galaxy's
envelope, it could not explain the galaxy's large number of globular clusters.
Thus, despite the susceptibility of dwarf galaxies to tidal disruption, there
is no strong evidence to support the idea that {\it most\/} of Virgo's
intracluster stars come from these objects.

On the other hand, there is evidence to suggest that most of Virgo's
intracluster stars come from non-dwarf galaxies.  The mean metallicity of the
Virgo RGB population lies in the range $-0.8 \lesssim$ [Fe/H] $\lesssim -0.2$;
the luminosity-metallicity relation of \citet{cote00} demonstrates that this is
substantially greater than that of most dwarfs.  It is, however, compatible
with models of galaxy harassment \citep{moore98,moore99}, in which intracluster
light is gradually built up by the removal of stars from sub-$L^*$ disks and
intermediate-luminosity E/S0 systems.  The high value of $\alpha$ derived above
for the intracluster population is also in broad agreement with these models.

It is important to note that this conclusion is by no means definitive. We
cannot rule out the existence of a substantial population of metal-poor stars,
so dwarf galaxies may still be an important contributor to Virgo's intracluster
light.  Furthermore, our results are still consistent with CDM models of cluster
formation.  Although metal-poor proto-galactic fragments do not appear to
dominate the intracluster light, most of these objects may have long ago been
assimilated into galaxies \citep[\eg][]{sz78,hp94}.   If so, then only the
small, relatively isolated systems would have lived long enough to form stars
and eventually be disrupted by the cluster's tidal field.  These stars could
easily be dominated by metal-rich objects liberated by galaxy harassment.

\section{Conclusions}

We have analyzed deep F814W {\sl HST\/} images of a single Virgo cluster field
located 41$^\prime$ NW of M87, near the cluster center.   Photometry of the
unresolved objects in this field \citep[combined with data from another Virgo
cluster field observed by][]{ftv98} shows an excess of objects (with respect to
the background HDF-N and HDF-S fields) with $I \gtrsim 27$, which we attribute
to intracluster RGB stars in the Virgo cluster.   We derive an average surface
brightness of $\mu_I = 27.9^{+0.3}_{-0.5}$ mag~arcsec$^{-2}$ for both fields;
if our data are representative of the cluster's IC light in general, then IC
stars comprise $15^{+7}_{-5}\%$ of Virgo's total light.   This result is
similar to that obtained from observations of IC planetary nebulae for values
of $\alpha_{2.5} = 23^{+10}_{-12} \times 10^{-9}$ PN $L_{\odot}^{-1}$.

We have modelled the resulting luminosity function with a single-component
RGB+AGB population, and derived the location of both the RGB tip ($I_{TRGB}=
27.31^{+0.27}_{-0.17}$) and the bright extent of an AGB ($\Delta I =
0.8^{+0.2}_{-0.2}$).  We note, however, that the latter is probably
contaminated by foreground RGB stars.  We find that the RGB tip is
significantly fainter than that observed in a Virgo cluster dE,N galaxy
\citep{har98}, and suggest that this difference is due to a higher metal
abundance for the intracluster stars ($-0.8\lesssim$ [Fe/H] $\lesssim -0.2$).
Our measurement of the intracluster AGB population indicates that the stars are
old ($t
> 2$~Gyr), but due to the possible existence of a foreground RGB component, we
cannot place a firm limit on the population age.  From our observations, it
seems most likely that the bulk of Virgo's intracluster stars were once
stripped from lower-mass spiral and elliptical galaxies, but we cannot rule out
the possibility that a significant metal-poor population (such as that expected
from tidally stripped {\it dwarf\/} galaxies) exists. It is clear that
measurements of the metallicity distribution of IC stars will be the key to
understanding their origins.

\acknowledgments

We would like to thank Peter Stetson for making his software and PSFs
available, and Leo Girardi for providing us with luminosity functions based on
his theoretical models.   Also thanks to Pat C\^{o}t\'{e}, Caryl Gronwall, Ken
Freeman and Magna Arnaboldi for related helpful discussions, and to Bruno
Binggeli for sending us his machine-readable version of the VCC.   We thank the
referee, Ted von Hippel, for his helpful comments that improved this paper.
This work has benefited greatly from use of facilities at the Canadian
Astronomy Data Centre, which is operated by the Herzberg Institute of
Astrophysics, National Research Council of Canada. This research has been
supported by grants NASA NAG5-9377 and HST-GO-08337.01-A.


\clearpage

\begin{deluxetable}{lcc}
\tablewidth{0pt} \tablecaption{Limiting Magnitudes} \tablehead{
\colhead{Field}&\colhead{$I_{lim}$} & \colhead{$\beta$} } \startdata
Virgo A &  27.69 & 1.55\\
FTV &  27.63 &
1.60\\
HDF-N &  27.61 & 1.62\\
HDF-S &  27.61 &
1.53\\
\enddata
\end{deluxetable}


\begin{deluxetable}{rcrrcrrcrr}
\tablewidth{0pt} \tablecaption{Luminosity Functions} \tablehead{ \colhead{} &
\colhead{}& \multicolumn{2}{c}{Virgo A\tablenotemark{a}} & \colhead{} &
\multicolumn{2}{c}{FTV\tablenotemark{a}} \\
\cline{3-4}
\cline{6-7}\\
\colhead{$I$} & \colhead{} &\colhead{$N_c$} & \colhead{$\sigma$}& \colhead{}
&\colhead{$N_c$} & \colhead{$\sigma$} \\
} \startdata
  25.5  &&   -1.2  &    1.5  &&    1.8   &   2.3\\
  25.6  &&    3.3  &    2.4  &&    3.3   &   2.4\\
  25.7  &&    3.5  &    2.1  &&    4.5   &   2.3\\
  25.8  &&   -1.2  &    1.5  &&    4.8   &   2.9\\
  25.9  &&    2.9  &    2.1  &&    8.9   &   3.3\\
  26.0  &&    1.1  &    2.7  &&    0.1   &   2.5\\
  26.1  &&    0.1  &    2.5  &&    3.1   &   3.0\\
  26.2  &&    0.5  &    3.0  &&    1.5   &   3.2\\
  26.3  &&    2.6  &    3.1  &&    3.6   &   3.2\\
  26.4  &&   10.6  &    4.2  &&    5.6   &   3.5\\
  26.5  &&   10.5  &    4.4  &&    3.5   &   3.5\\
  26.6  &&    5.5  &    3.8  &&   10.5   &   4.4\\
  26.7  &&   14.2  &    5.1  &&   14.2   &   5.1\\
  26.8  &&   23.8  &    6.5  &&   17.8   &   6.1\\
  26.9  &&   14.4  &    6.3  &&   24.4   &   7.1\\
  27.0  &&   17.1  &    6.9  &&   24.1   &   7.4\\
  27.1  &&   43.6  &    8.6  &&   29.6   &   7.7\\
  27.2  &&   51.8  &    9.8  &&   31.8   &   8.7\\
  27.3  &&   81.2  &   11.2  &&   47.2   &   9.6\\
  27.4  &&  106.1  &   12.4  &&   68.1   &  10.7\\
  27.5  &&  130.8  &   13.2  &&   78.8   &  11.1\\
  27.6  &&  129.8  &   13.2  &&   71.8   &  10.8\\
\enddata
\tablenotetext{a}{background-corrected}
\end{deluxetable}

\clearpage

\begin{deluxetable}{lcccc}
\tablewidth{0pt} \tablecaption{Luminosity Functions - Best Fit Models}
\tablehead{ \colhead{Field}&\colhead{$I_{TRGB}$} & \colhead{$\Delta I_{AGB}$}
&\colhead{$\mu_{I}(RGB+AGB)$\tablenotemark{1}} &\colhead{$\chi^2_{\nu}$}}
\startdata
Virgo A &  $27.31^{+0.39}_{-0.19}$ & $0.7^{+0.3}_{-0.3}$ & $30.08^{+0.10}_{-0.18}$ & 1.14\\
FTV &  $27.51^{+0.34}_{-0.72}$  & $0.9^{+0.2}_{-0.3}$  & $30.42^{+0.26}_{-0.09}$ & 0.48 \\
Combined & $27.31^{+0.27}_{-0.17}$  & $0.8^{+0.2}_{-0.2}$ & $30.28^{+0.11}_{-0.08}$ & 0.89 \\
\enddata
\tablenotetext{1}{$I$-band surface brightness of scaled LF model for RGB and
AGB stars brighter than $I=28.1$}
\end{deluxetable}


\begin{deluxetable}{ccc}
\tablewidth{0pt} \tablecaption{Missing Light Corrections - $I$ filter}
\tablehead{ \colhead{$\Delta
I_{RGB}$\tablenotemark{a}}&\colhead{$F$\tablenotemark{b}} & \colhead{$\Delta
\mu_{ML}$\tablenotemark{c}} } \startdata
0.3 & $0.047\pm 0.006$ & $-3.33 \pm 0.14$ \\
0.4 & $0.061\pm
0.006$ & $-3.03 \pm 0.11$ \\
0.5 & $0.074\pm 0.007$ & $-2.82 \pm 0.11$ \\
0.6 &
$0.088\pm 0.008$ & $-2.64 \pm 0.10$ \\
0.7 & $0.100\pm 0.008$ & $-2.50 \pm 0.09$
\\
0.8 & $0.113\pm 0.009$ & $-2.37 \pm 0.09$ \\
0.9 & $0.124\pm 0.010$ & $-2.26 \pm
0.08$ \\
1.0 & $0.136\pm 0.010$ & $-2.17 \pm 0.08$
\\
\enddata
\tablenotetext{a}{$\Delta I_{RGB} = I_{cut} - I_{TRGB} = 28.0 - I_{TRGB}$}
\tablenotetext{b}{mean fraction of total luminosity (with no AGB) within
$\Delta I_{TRGB}$, with rms error} \tablenotetext{c}{`missing light'
correction}
\end{deluxetable}

\clearpage

\begin{figure}
\plotone{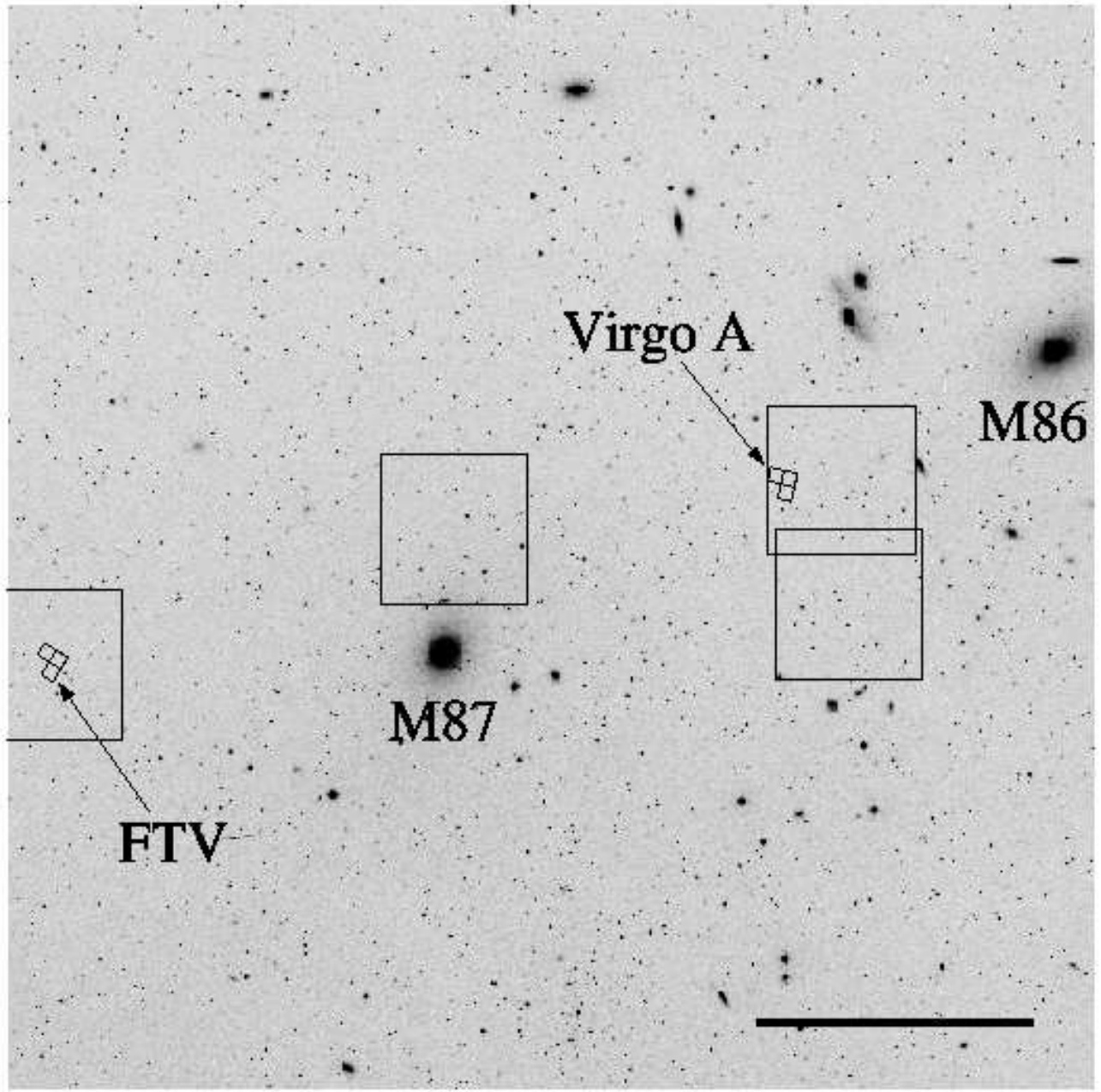} \caption{A Digitized Sky Survey image of the central region of
the Virgo Cluster, with the location of our Subclump A field and the FTV survey
field superposed.   The image is 2$^\circ$ on a side, with north at the top and
east to the left. The solid line at the lower right represents $30^\prime$.
Also shown are four \citet{feld00} survey fields for intracluster planetary
nebulae. \label{fig1}}
\end{figure}

\begin{figure}
\plotone{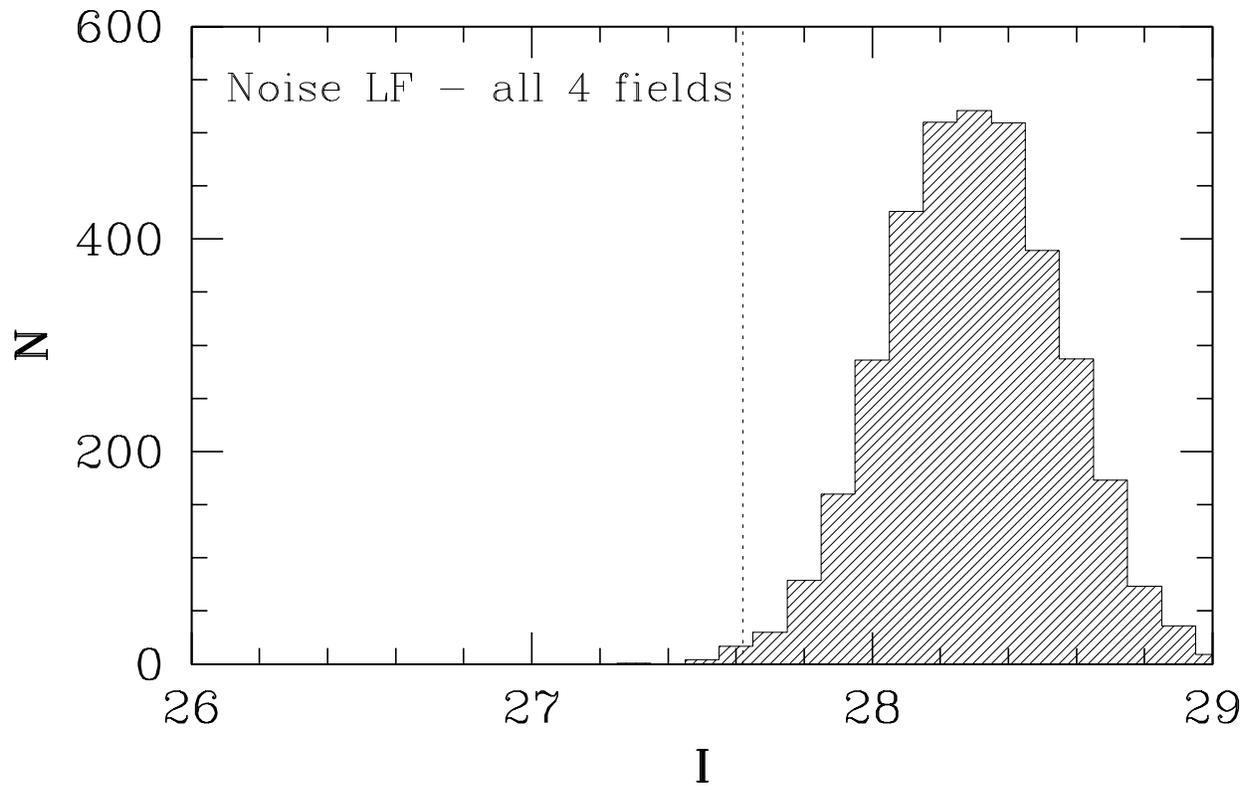}\caption{The ``luminosity function'' of noise spikes detected
in the inverse images of all four {\sl HST\/} fields. The data have been binned
into 0.1~mag intervals. The dotted line shows the limiting magnitude of our
survey. Because of our low detection threshold, the number of false detections
is significant for magnitudes $I > 27.6$. At brighter magnitudes, however,
contamination of the luminosity by false detections is not important.
\label{fig2}}
\end{figure}

\begin{figure} \plotone{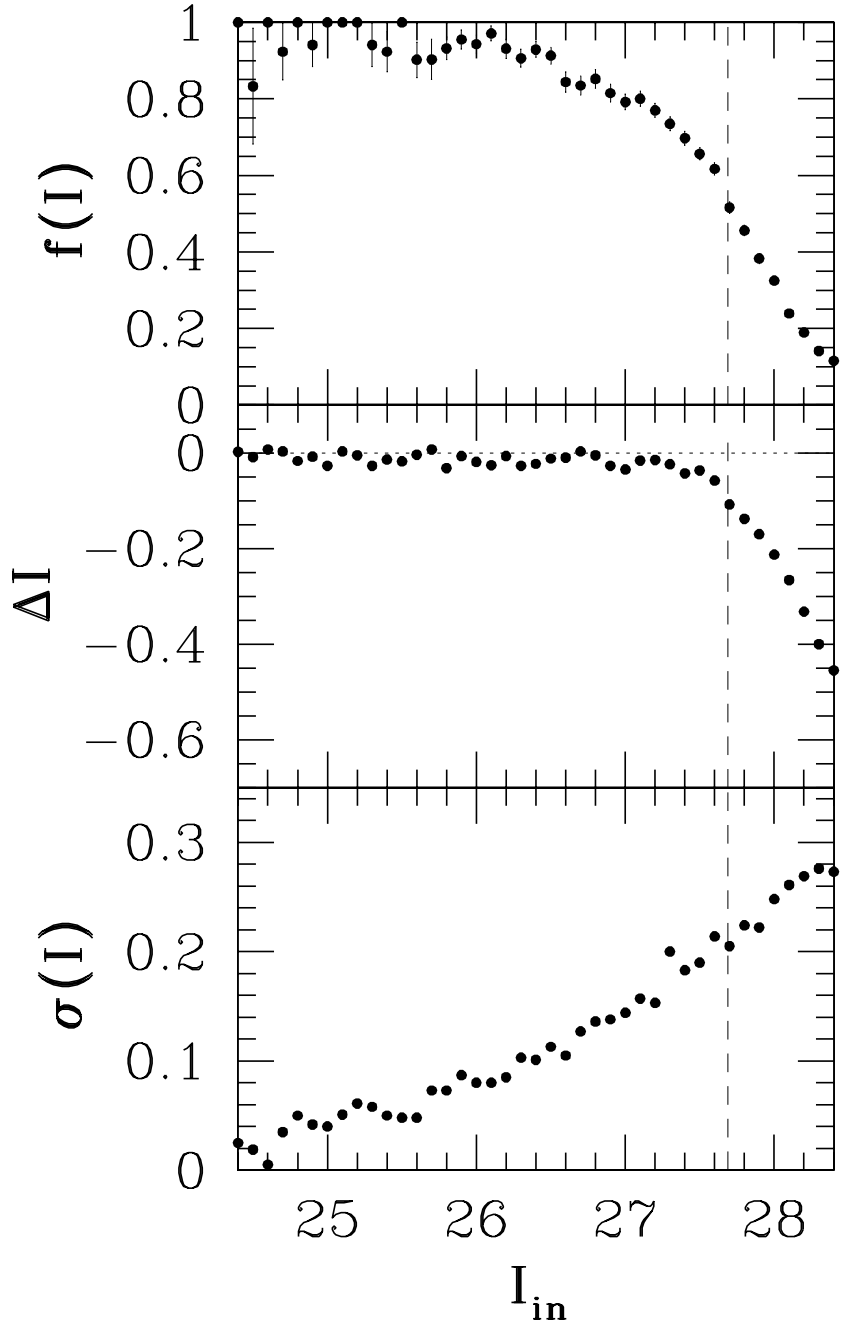}\caption{Results from artificial star experiments
performed on the three WF images in our Virgo A field. The data have been
binned into 0.1~mag intervals; $I_{in}$ is the input magnitude of the added
stars. The top panel plots $f(I)$, the fraction of artificial stars recovered
by our detection algorithm.  The center and bottom panels show how the measured
magnitudes of the recovered stars relate to the input magnitudes: the center
panel plots the difference between the mean measured magnitude and the input
magnitude, while the bottom panel gives the dispersion of the distribution. The
limiting magnitude of our survey (shown by the dotted line), is defined as the
place where the fraction of objects recovered drops to 50\%.\label{fig3}}
\end{figure}

\begin{figure}
\plotone{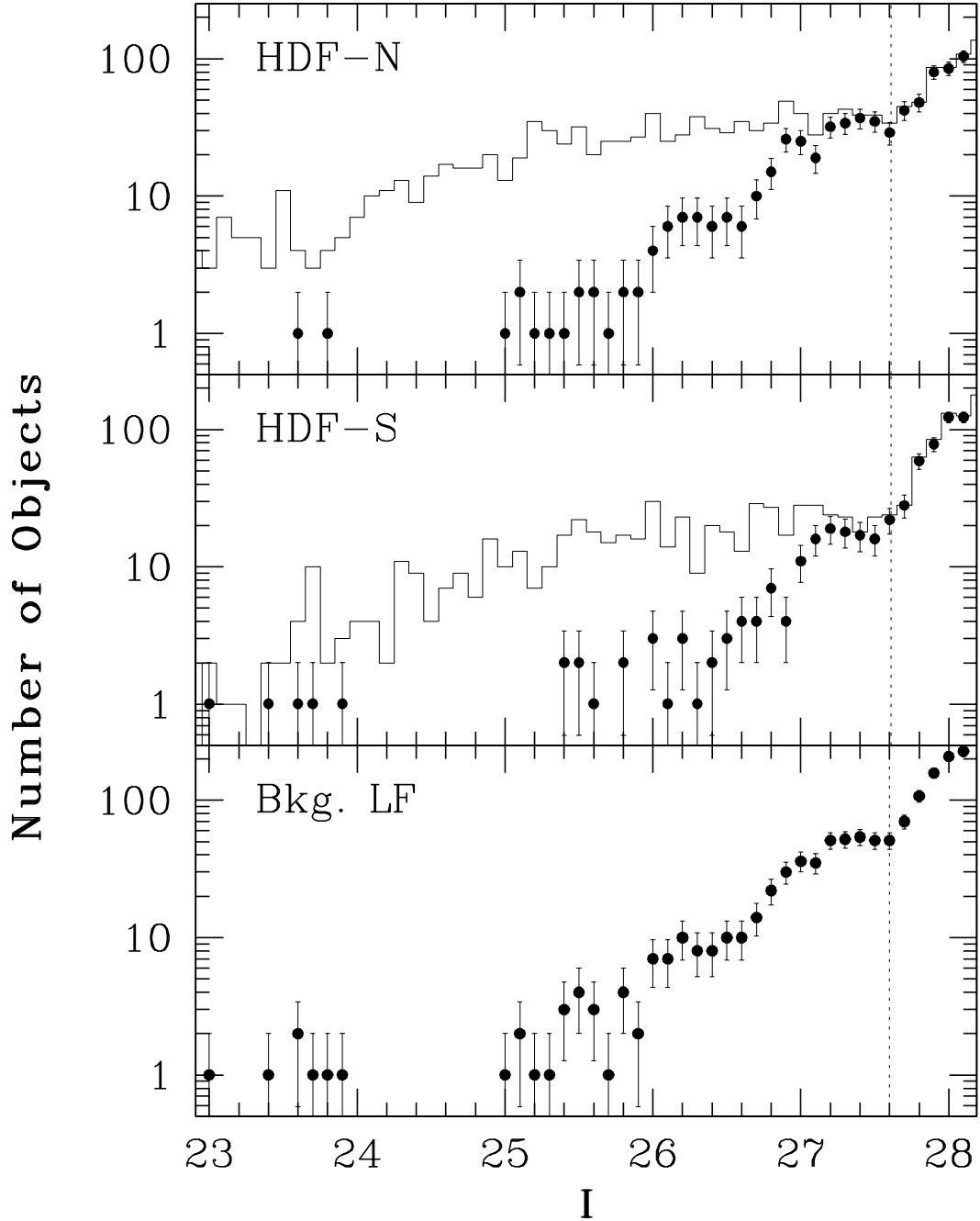}\caption{The $I$-band luminosity functions (LFs) of the HDF-N
and HDF-S fields, binned into 0.1~mag intervals.  The histograms give the total
source counts; the solid circles, with their Poissonian error bars, show the
LFs after the removal of non-stellar sources.  Bins with no counts are not
displayed. The dotted line notes $I_{lim}$, the magnitude where the data is
50\% complete. When scaled by their survey areas, the LFs of HDF-N and HDF-S
are statistically indistinguishable; their counts can therefore be added to
create a total `background' LF{}. This is displayed in the bottom panel of the
figure.\label{fig4}}
\end{figure}

\begin{figure}
\plotone{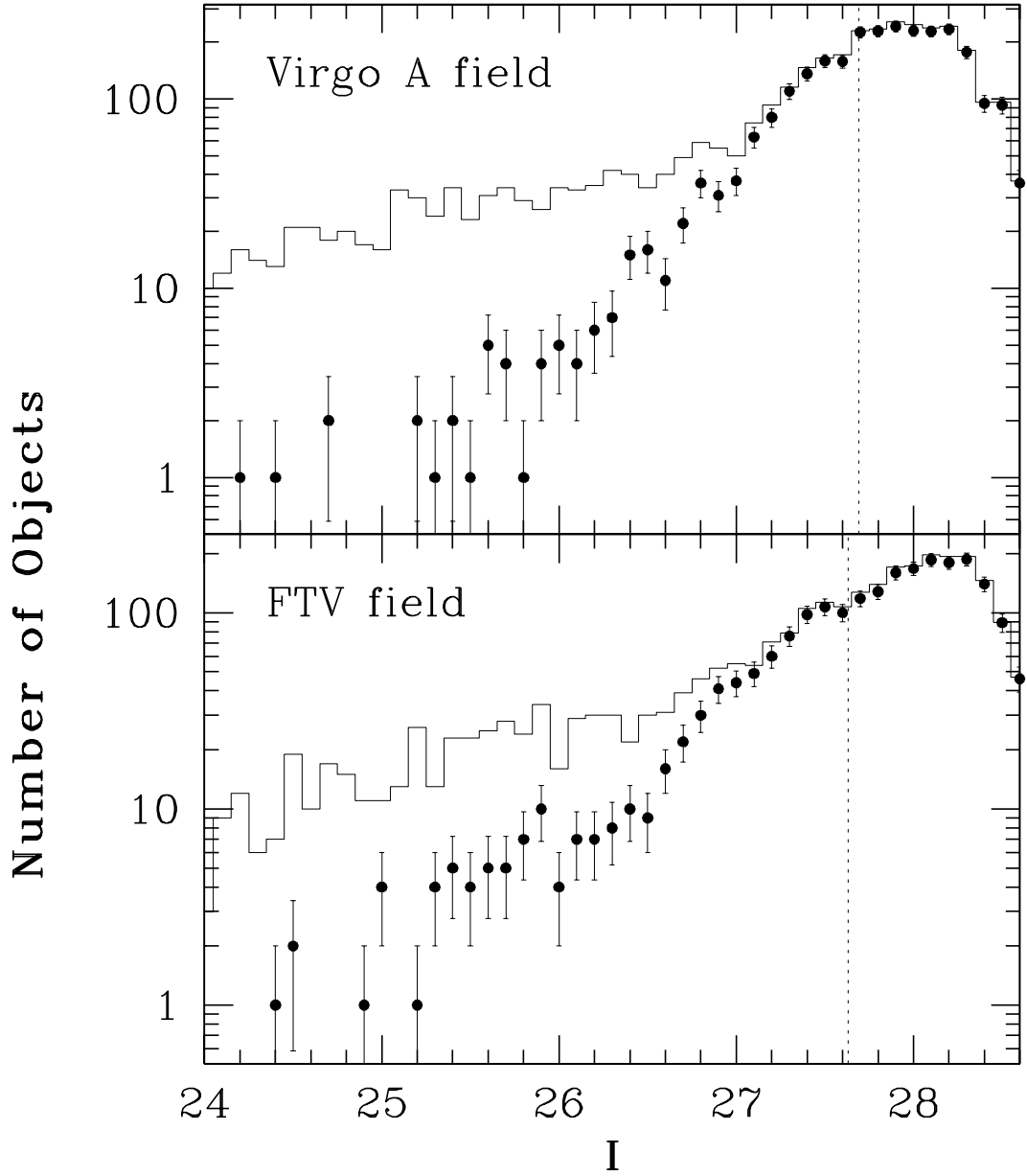}\caption{The $I$-band luminosity functions for the Virgo~A
field and the FTV field, binned into 0.1~mag intervals. The histograms give the
total source counts; the solid circles, with their Poissonian error bars, show
the LFs after the removal of non-stellar sources.  The dotted line notes
$I_{lim}$, the magnitude where the data is 50\% complete.  As with Figure 4,
bins with no source counts are not plotted. \label{fig5}}
\end{figure}

\begin{figure}
\plotone{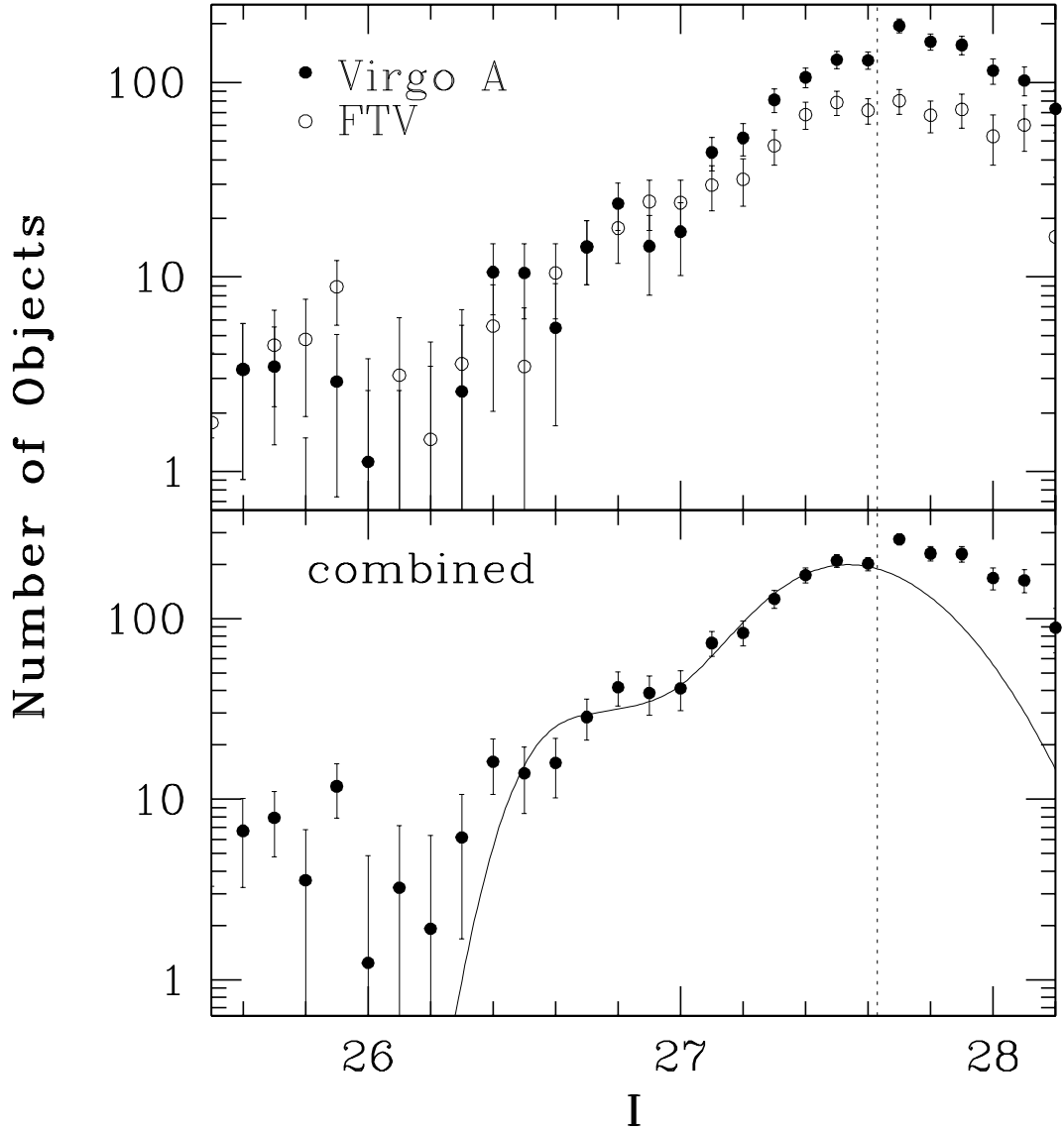}\caption{The $I$-band luminosity function (LF) for point
sources in our Virgo A field and in the FTV field with the LF of the Hubble
Deep Fields removed. The data have been binned into 0.1~mag intervals, and the
error bars reflect the Poissonian uncertainties of the Virgo and HDF fields
added in quadrature. The data demonstrate that Virgo contains an excess of
point sources with $I \gtrsim 26.4$ that this excess becomes dramatically
larger at $I \gtrsim 26.8$. The lower panel shows the combination of both
background-subtracted LFs, and the corresponding best-fitting RGB+AGB model
(see text for details) \label{fig6}}
\end{figure}

\begin{figure}
\plotone{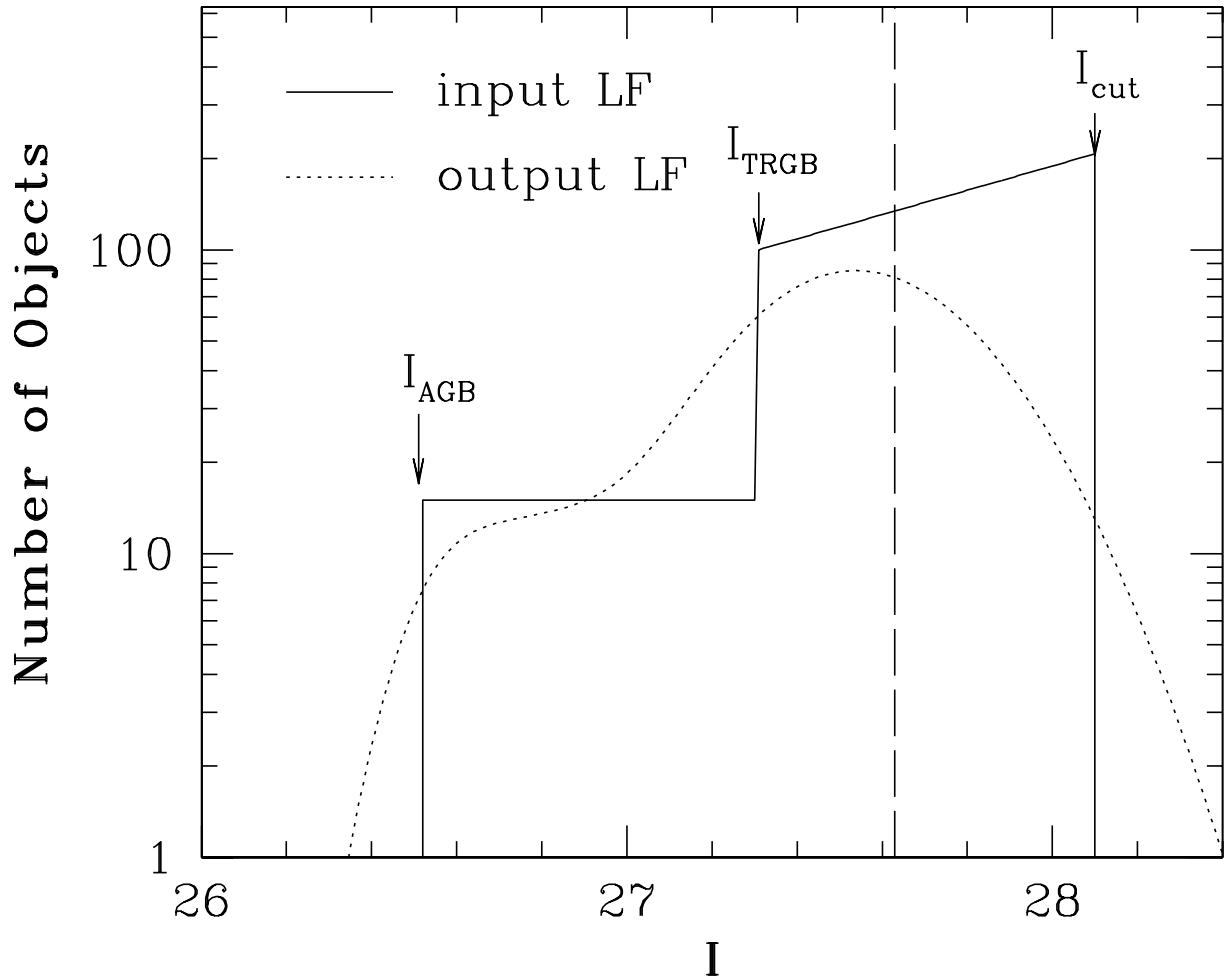}\caption{A schematic of the intrinsic RGB $+$ AGB luminosity
function for Virgo's intracluster stars.  The AGB to RGB normalization at the
tip of the red giant branch is fixed at 15\%; therefore this luminosity
function has three free parameters:  the tip of the RGB ($I_{TRGB}$), the
magnitude difference between the AGB and RGB tip ($\Delta I_{AGB} = I_{TRGB} -
I_{TAGB}$), and the overall normalization of the function. The smooth (dashed)
curve is our best-fitting observed LF, which is the intrinsic LF (solid line)
convolved with the photometric error function and corrected for incompleteness.
\label{fig7}}
\end{figure}

\begin{figure}
\plotone{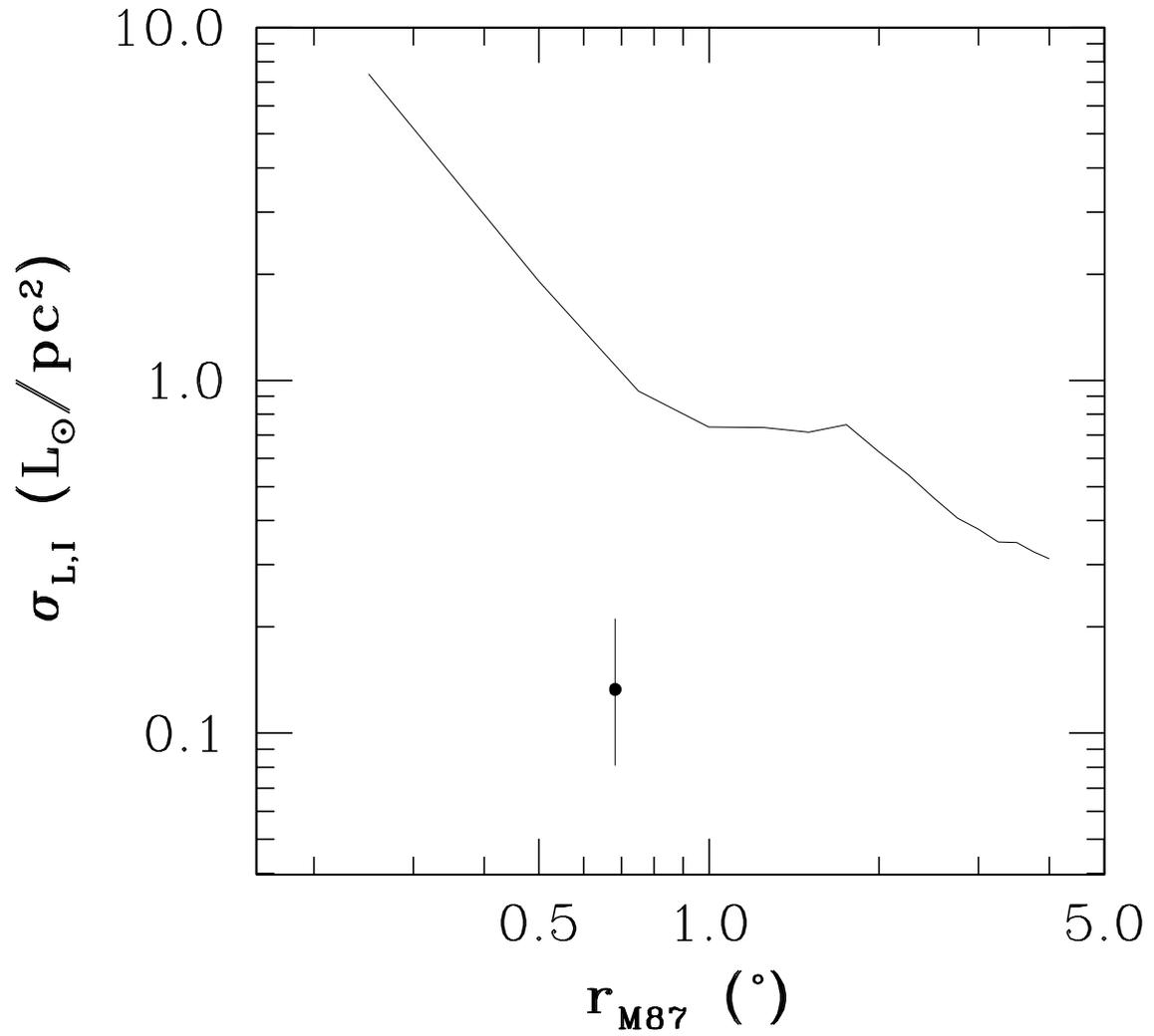}\caption{The cumulative luminosity surface density
$\sigma_{L,I}$ (solid line) for member galaxies in the Virgo cluster, as a
function of radial distance from M87.  The filled circle denotes the surface
brightness of intracluster stars derived from our best-fitting model, and
includes the corresponding 1-$\sigma$ error bars.\label{fig8}}
\end{figure}

\end{document}